\documentclass[usenatbib]{mn2e}
\usepackage{graphicx, amssymb, aas_macros, natbib}
\newcommand{\vmax}{V_{\rm{max}}}
\newcommand{\rmax}{R_{\rm{max}}}
\newcommand{\mhalf}{M_{1/2}}
\newcommand{\rhalf}{R_{1/2}}

\newcommand{\msub}{M_{\rm{sub}}}

\newcommand{\mstar}{M_{\star}}
\newcommand{\msun}{M_{\odot}}

\newcommand{\lsun}{L_{\odot}}

\newcommand{\mpc}{{\rm Mpc}}

\newcommand{\kms}{{\rm km \, s}^{-1}}

\newcommand{\vacc}{V_{\rm infall}}

\newcommand{\lcdm}{$\Lambda$CDM}

\newcommand{\flux}{\mathcal{F}}

\title[Massive dark subhaloes in the Milky Way]
{
Too big to fail?  The puzzling darkness of massive Milky Way subhaloes
}
\author[M. Boylan-Kolchin, J. S. Bullock, and M. Kaplinghat]
{Michael Boylan-Kolchin\thanks{$\!$Center for Galaxy Evolution
  fellow}\thanks{email: m.bk@uci.edu}, James S. Bullock, and Manoj Kaplinghat\\
\noindent $\!\!$Center for Cosmology, Department of Physics and Astronomy, 
  4129 Reines Hall, University of California, Irvine, CA 92697, USA}
\begin{document}

 \pagerange{\pageref{firstpage}--\pageref{lastpage}} 
 \pubyear{2011}

\maketitle

\label{firstpage}
\begin{abstract}
  We show that dissipationless \lcdm\ simulations predict that the majority of
  the most massive subhaloes of the Milky Way are too dense to host any of its
  bright satellites ($L_V > 10^{5}\,\lsun$).  These dark subhaloes have peak
  circular velocities at infall of $\vacc = 30-70\,\kms$ and infall masses of
  $[0.2-4]\times 10^{10}\,\msun$. Unless the Milky Way is a statistical anomaly,
  this implies that galaxy formation becomes effectively stochastic at these
  masses. This is in marked contrast to the well-established monotonic relation
  between galaxy luminosity and halo circular velocity (or halo mass) for more
  massive haloes. We show that at least two (and typically four) of these massive
  dark subhaloes are expected to produce a larger dark matter annihilation flux
  than Draco. It may be possible to circumvent these conclusions if baryonic
  feedback in dwarf satellites or different dark matter physics can reduce the
  central densities of massive subhaloes by order unity on a scale of 0.3 -- 1
  kpc.
\end{abstract}


\section{Introduction} 
The cold dark matter (CDM) paradigm has been demonstrably successful at
explaining a variety of observations on cosmological scales.  Tests on smaller
scales are often complicated by the physics of galaxy formation, but are crucial
for verifying the CDM model.  Perhaps the most prominent issue facing \lcdm\ on
galactic scales is the large discrepancy between the number of observed and
expected satellite galaxies of the Milky Way (\citealt{kauffmann1993,
  klypin1999, moore1999}; see \citealt{bullock2010a} for a recent review).
Accordingly, much theoretical work has been devoted to understanding how to
reproduce the satellite population of the Milky Way (MW).

While some \lcdm\ models of the MW's satellite population place the most
luminous dwarf galaxies in the most massive subhaloes at redshift zero
\citep{stoehr2002, hayashi2003, penarrubia2008}, recent kinematic studies of the
satellites have shown that this is unlikely to be the case \citep{walker2009,
  strigari2010}.  Other models postulate that MW satellite galaxies correspond
to subhaloes that were the most massive at some earlier time \citep{bullock2000,
  kravtsov2004, ricotti2005, koposov2009, okamoto2009, busha2010}, often the
epoch of reionization, with galaxy formation strongly suppressed in lower mass
subhaloes (see \citealt{kravtsov2010} for a recent review).  In addition, many
faint MW satellites have been discovered in the SDSS (e.g.,
\citealt{willman2005, belokurov2007}), and it has become clear that up to a
factor of $\sim 5-20$ times as many faint galaxies could remain undetected at
present owing to incomplete sky coverage, luminosity bias, and surface brightness
limits \citep{tollerud2008, walsh2009, bullock2010}.

While this theoretical and observational progress has alleviated -- though not
eliminated -- concerns about the mismatch between the number of low-mass
subhaloes and faint MW satellites, a pressing question remains: 
{\it what is the stellar content of massive Milky Way subhaloes at redshift zero?}
In this Letter, we focus on properties of massive subhaloes in \lcdm\ galaxy-mass
haloes, and examine which Milky Way satellites -- if any -- can be hosted by such
subhaloes.  

\section{Simulations and Data}
\label{section:methods}
Our \lcdm\ predictions are based on dark matter subhaloes from the Aquarius
project \citep{springel2008} and the Via Lactea II simulation
\citep{diemand2008, diemand2007a}.  The Aquarius project consists of six
galaxy-mass haloes -- denoted A through F -- simulated at a series of
increasingly high mass and force resolution.  Although only halo A was simulated
at the highest resolution level, all six haloes were simulated with particle mass
$m_p = (0.64-1.4) \times 10^{4}\,\msun$ and Plummer-equivalent
gravitational softening $\epsilon = 66$ pc; it is this set of ``level 2''
simulations that we use in this paper.  The Via Lactea II simulation (VL-II) of
one galaxy-mass halo used $m_p=4.1\times 10^{3}\,\msun$ and $\epsilon=40$ pc.  A
notable difference between the simulations is the background cosmological model:
the Aquarius simulations used a value of 0.9 for the power spectrum
normalization $\sigma_8$ and 1.0 for the spectral index of the primordial power
spectrum $n_s$, while VL-II used $\sigma_8=0.74$ and $n_s=0.951$.  The best
current estimates of these parameters, $\sigma_8=0.816 \pm 0.024$ and $n_s=0.968
\pm 0.012$ \citep{komatsu2011}, fall in between those used for the simulations.

In each simulation, we select every subhalo that lies within 300 kpc of the
host's center and has a maximum circular velocity $\vmax \equiv {\rm max}\{[G
M(<\!R)/R]^{1/2}\}$ exceeding $10 \,\kms$.  We characterize a subhalo prior to
infall onto its host via $\vacc$, which we define to be the value of $\vmax$
when the subhalo's mass was at a maximum (over its entire evolution) in Aquarius
and the maximum value of $\vmax$ over the subhalo's entire history for VL-II.
The measured values of $\vmax$ and $\rmax$ (the radius at which $\vmax$ is
attained) at redshift zero are used to determine each subhalo's inner mass
distribution by assuming that the subhalo's density structure can be modeled by
a \citet*[hereafter NFW]{navarro1997} profile with the same $\vmax$ and $\rmax$.
Using subhaloes extracted from the Millennium-II Simulation
\citep{boylan-kolchin2009}, we have verified that this approach gives the
correct mass to better than 10\% at radii that are well resolved\footnote{While
  both the host haloes and subhaloes from Aquarius are fit somewhat better by
  \citet{einasto1965} profiles than by NFW profiles \citep{navarro2010,
    springel2008}, we use NFW profiles here because they provide more
  conservative constraints: at fixed $\vmax$ and $\rmax$, an Einasto profile
  contains more mass than an NFW profile within a given radius $R$ for
  reasonable values of the Einasto shape parameter $\alpha$ when $R < \rmax$.}
(as expected from earlier work by \citealt{kazantzidis2004b}).

To connect the $N$-body subhaloes to the bright ($L_V > 10^{5}\,\lsun$) dwarf
spheroidal galaxies of the Milky Way, we turn to kinematic measurements of the
dwarfs' masses.  \citet{walker2009} and \citet{wolf2010} have recently shown
that dispersion-supported galaxies such as the MW dwarf spheroidals have
dynamical masses $\mhalf$ within their deprojected half-light radii $\rhalf$
that are well-constrained by line-of-sight velocity measurements.  Since these
galaxies are all strongly dark matter-dominated even within $\rhalf$ (e.g.,
\citealt{mateo1998}), observed values of $\mhalf$ are effectively measurements
of the dark matter mass within $\rhalf$.  A necessary, but not sufficient,
condition for a subhalo to possibly host a given dwarf is that $\msub(\rhalf)$
agree with $M_{\rm dwarf}(\rhalf)$.  Conversely, a dwarf cannot live in a
subhalo if $\msub(\rhalf)$ and $M_{\rm dwarf}(\rhalf)$ differ substantially.

\begin{figure}
 \centering
 \includegraphics[scale=0.54, viewport=20 0 400 420]{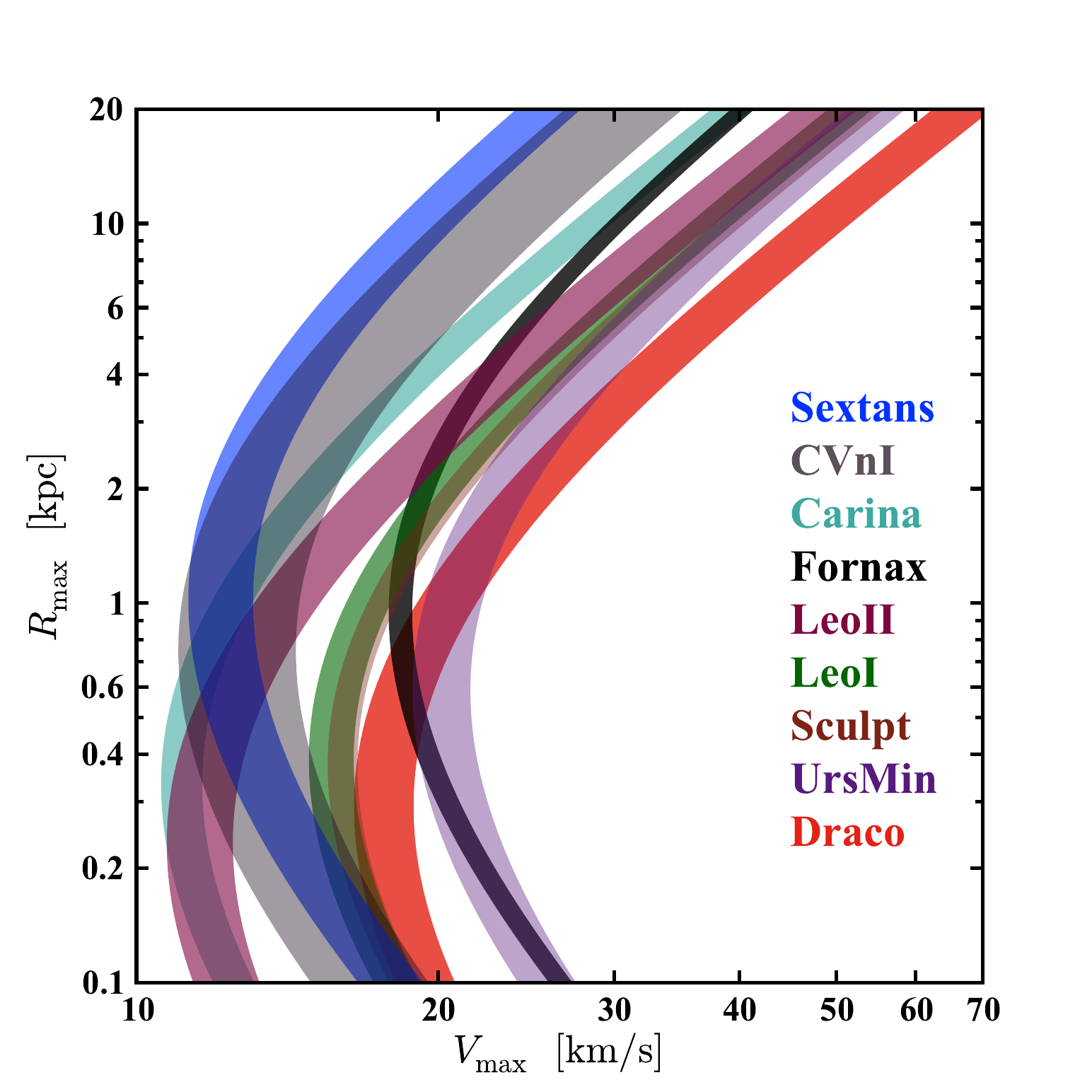}
 \caption{Constraints on the $\vmax-\rmax$ values (assuming NFW profiles) of the
   hosts of the nine bright ($L_V > 10^5\,\lsun$) MW
   dwarf spheroidal galaxies.  The colored bands show $1\,\sigma$ confidence
   intervals based on measured values of $\rhalf$ and $\mhalf$ from \citet{wolf2010}.
 \label{fig:dwarfs}
}
\end{figure}

Given the values of $\mhalf$ calculated by \citet{wolf2010}, we can therefore
investigate what $\{\rmax, \vmax\}$ values of NFW subhaloes are consistent with
the observed dynamics of the bright MW dwarf spheroidals.  We exclude
Sagittarius, which is far from dynamical equilibrium, for the present.
Figure~\ref{fig:dwarfs} shows the resulting $1\,\sigma$ confidence regions in
$\vmax$-$\rmax$ space for these nine dwarfs.  The behavior of the contours for
each of the dwarfs is qualitatively similar: there is a global minimum in
$\vmax$, corresponding to $\rmax = \rhalf$ and $\vmax=\sqrt{3}\,\sigma_{\rm los,
  \star}$ \citep{wolf2010}, with allowed values of $\vmax$ increasing for both
smaller and larger values of $\rmax$ (corresponding to $\rmax < \rhalf$ and
$\rmax > \rhalf$, respectively).  The lower portions of the curves, where $\rmax
< \rhalf$, are unlikely to be physically plausible models for the hosts of
dwarfs, as they require that the dark matter subhalo has been strongly affected
by tides on the scale of the luminous matter in the dwarf.

\section{Results}
\label{section:results}
\begin{figure}
 \centering
 \includegraphics[scale=0.54, viewport=28 0 510 420]{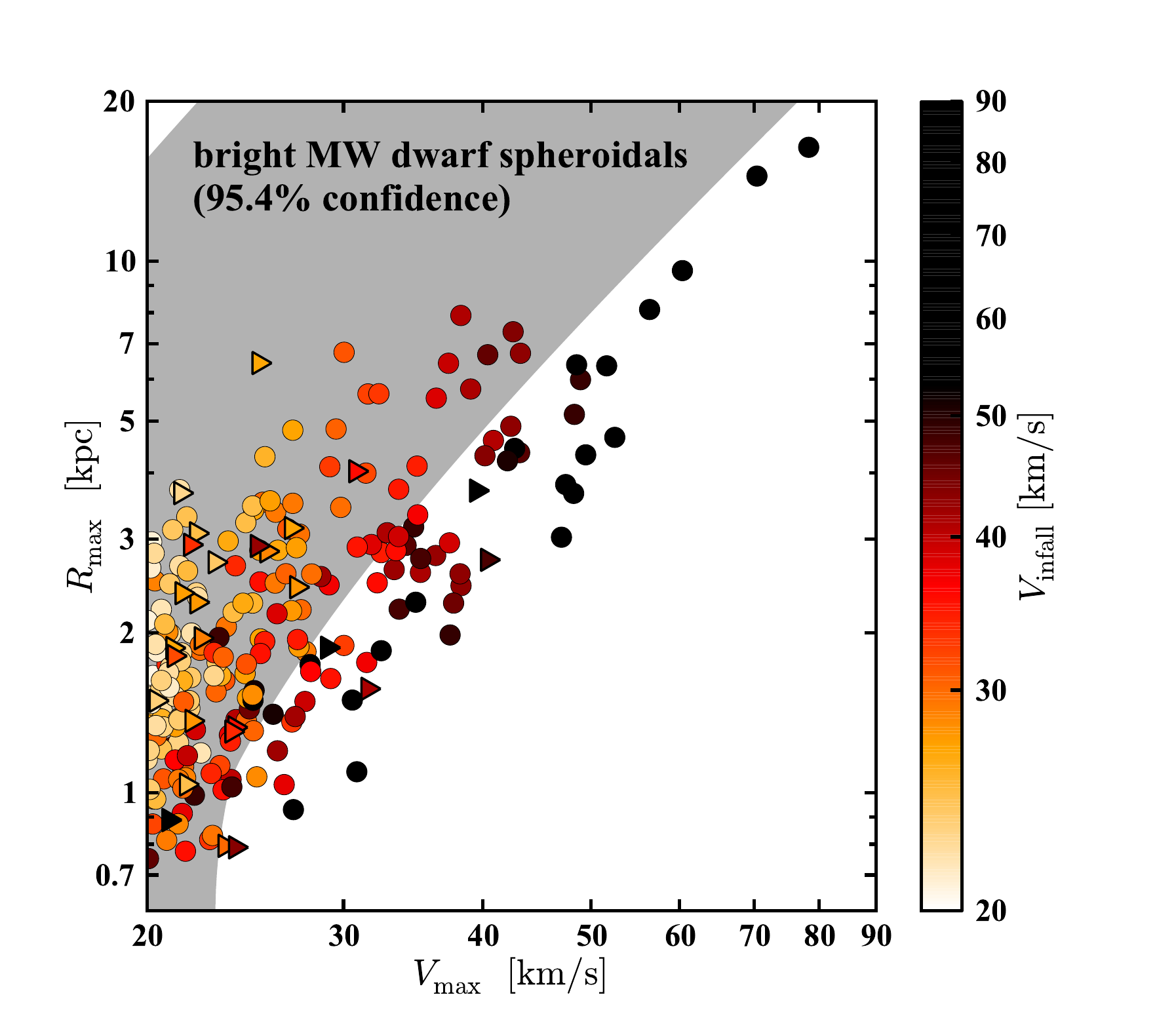}
 \caption{Subhaloes from all six Aquarius simulations (circles) and Via Lactea II
   (triangles), color-coded according to $\vacc$. The shaded gray
   region shows the $2\,\sigma$ confidence interval for possible hosts of the
   bright MW dwarf spheroidals (see Fig.~\ref{fig:dwarfs}).
   \label{fig:rmax_vmax}
 }
\end{figure}

In Figure~\ref{fig:rmax_vmax}, we plot data for all subhaloes from the six
Aquarius simulations (circles) and from the VL-II simulation (triangles),
color-coded by $\vacc$.  The gray shaded band corresponds to $2\,\sigma$
constraints from the MW dwarf spheroidal galaxies in Fig.~\ref{fig:dwarfs}.  In
terms of the total mass within 300 parsecs ($M_{300}$; \citealt{strigari2008}),
this gray shaded region is almost exactly the same as $6.5\times 10^{6} <
M_{\rm 300} /\msun < 3\times 10^{7}$.  Many of the subhaloes lie in the range
that is consistent at the $2\,\sigma$ level with the dwarfs, but there is a
large population of subhaloes that does not.  These subhaloes all have central
densities that are too high to host any of the bright MW dwarf spheroidals; they
also have higher values of both $\vmax$ and $\vacc$, on average.

The Milky Way contains three additional satellites that are brighter than the
nine dwarf spheroidals included in Figs.~\ref{fig:dwarfs}-\ref{fig:rmax_vmax}:
the Large Magellanic Cloud (LMC), the Small Magellanic Cloud (SMC), and the
Sagittarius dwarf spheroidal.  In the context of \lcdm\ models of galaxy
formation, the Magellanic Clouds are expected to reside in subhaloes with large
values of $\vacc$: using abundance matching (e.g., \citealt{kravtsov2004a,
  conroy2006, guo2010}) to assign stellar mass to subhaloes\footnote{We match
  $n(>\mstar)$ from \citet{li2009} to $n(>\vacc)$ that we have calculated from
  the Millennium and Millennium-II simulations \citep{springel2005a,
    boylan-kolchin2009}.}, we find that the SMC should have $\vacc = 70-80
\,\kms$ and the LMC should have $\vacc = 95-105 \,\kms$.  Conservatively, we
estimate that the Magellanic Clouds have $\vacc > 60\,\kms$ and $\vmax(z=0) >
40\,\kms$ \citep{stanimirovic2004, olsen2007} and remove all subhaloes with these
properties from our sample of subhaloes that are inconsistent with the dynamics
of the bright MW dwarfs.  Sagittarius is in the process of being completely
disrupted by the disk, but estimates of its pre-interaction properties give it a
total stellar mass similar to the SMC \citep{niederste-ostholt2010}.  In the
absence of the MW disk -- e.g., in dissipationless simulations such as those
used here -- it is very likely that Sagittarius would be much more massive at
$z=0$, and our Magellanic Cloud exclusion criteria might also be appropriate for
Sagittarius.  Regardless, the inclusion or exclusion of one object does not
alter the conclusions reached below. The remaining subhaloes are not compatible
with hosting any of the bright ($L_V > 10^5\,\lsun$) satellites of the Milky
Way; we refer to these as massive dark subhaloes and focus the remainder of our
analysis on them.

\begin{figure}
 \centering
 \includegraphics[scale=0.54, viewport=20 0 400 420]{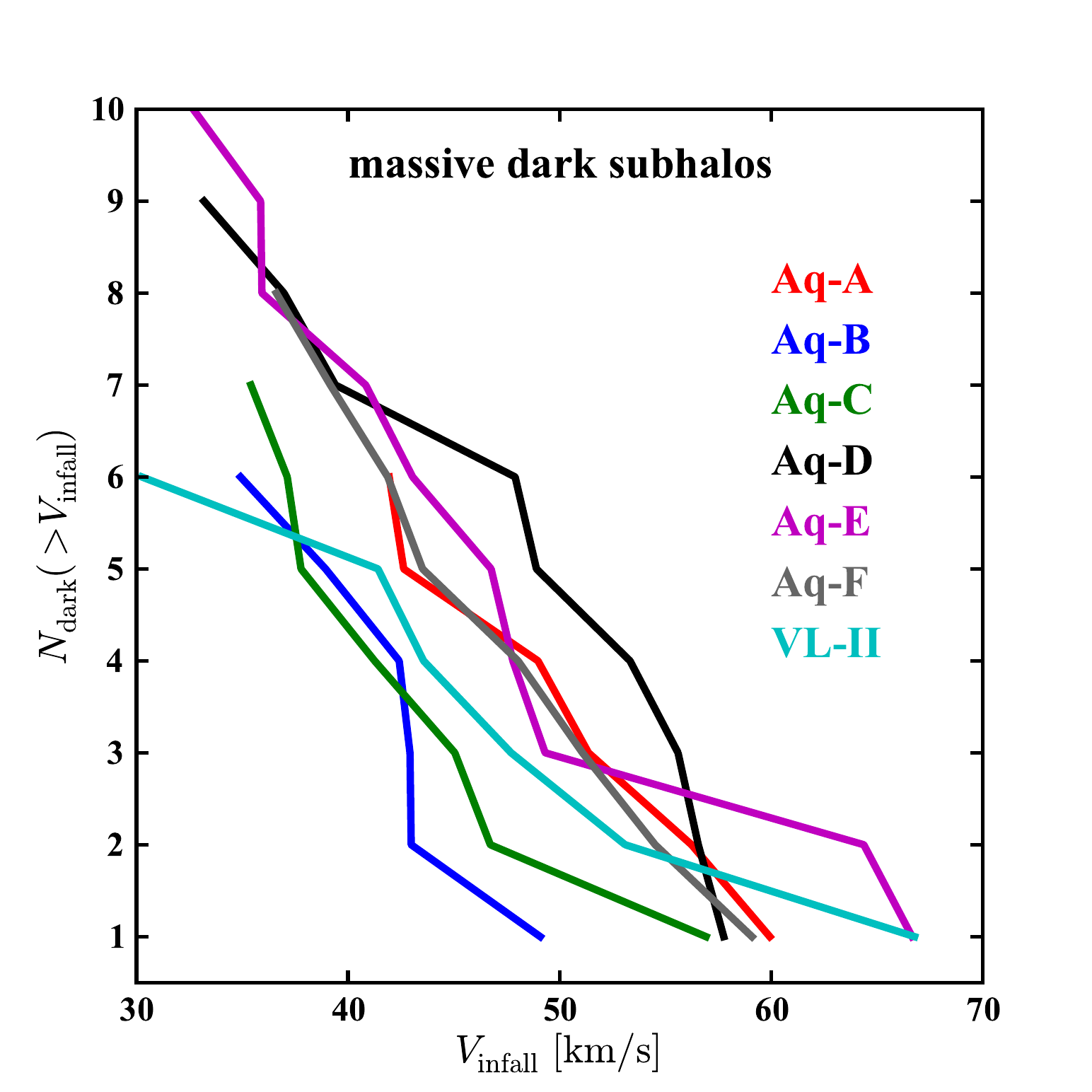}
 \caption{Cumulative $\vacc$ function of massive subhaloes 
   at $z=0$ that cannot host any MW satellite brighter than
   $L_V=10^{5}\,\lsun$, including the Magellanic Clouds.  Each of the seven high
   resolution simulations studied here has at least six such subhaloes with
   $\vacc > 30\,\kms$, and at least four with $\vacc > 40\,\kms$.  
   \label{fig:vacc_cum}
 }
\end{figure}

Figure~\ref{fig:vacc_cum} shows the cumulative velocity function of massive dark
subhaloes as a function of $\vacc$ for each of the seven simulations considered
here.  All of the dark subhaloes plotted in Fig.~\ref{fig:vacc_cum} have current
$\vmax$ values larger than 23 $\kms$, and none meet our criteria for hosting
galaxies similar to the Magellanic Clouds.  In all cases, there are at least 6
-- and up to 12 -- subhaloes with $\vacc > 30\,\kms$ that are not consistent with
any of the bright MW satellites (i.e, any satellite with $L_V > 10^{5}\,\lsun$).
These subhaloes tend to be more massive than the possible hosts of the dwarf
spheroidals, both today and at infall (see Fig.~\ref{fig:sham} below).
Moreover, the three haloes with the fewest massive dark subhaloes (Aq-B, Aq-C, and
VL-II) do not contain any potential Magellanic Cloud hosts.  If we restrict
ourselves to the simulations that do contain reasonable Magellanic Cloud
analogs, then the predicted number of massive dark subhaloes is closer to 10,
including several with $\vacc > 50\,\kms$.

The luminosity - $\vacc$ relation for a representative halo is shown in
Figure~\ref{fig:sham}.  The MW dwarfs (red symbols) were assigned their $\vacc$
values by placing the most luminous dwarf spheroidal (Fornax) in the subhalo
with the largest value of $\vacc$ that has $\msub(\rhalf)$ within $1\,\sigma$ of
the measured $\mhalf$ of Fornax, then repeating the process for each of the
other dwarfs in order of decreasing luminosity.  For each dwarf, the assigned
value of $\vacc$ can therefore be considered an upper limit at $68\%$ confidence
within this realization.  The massive dark subhaloes (black symbols) are placed
on the same plot according to their $\vacc$.  These subhaloes must all have
luminosities less than $10^{5}\,L_{\odot}$ in order to have escaped detection in
all-sky optical surveys \citep{whiting2007}.  The dotted blue line shows an
extrapolation of abundance matching, assuming $\mstar/L_V=1$.  It is clear that
neither the bright dwarf spheroidals nor the dark subhaloes described in this
paper can be easily accommodated by galaxy formation models in which luminosity
is a monotonic function of halo mass or $\vacc$.

\begin{figure}
 \centering
 \includegraphics[scale=0.54, viewport=30 0 400 420]{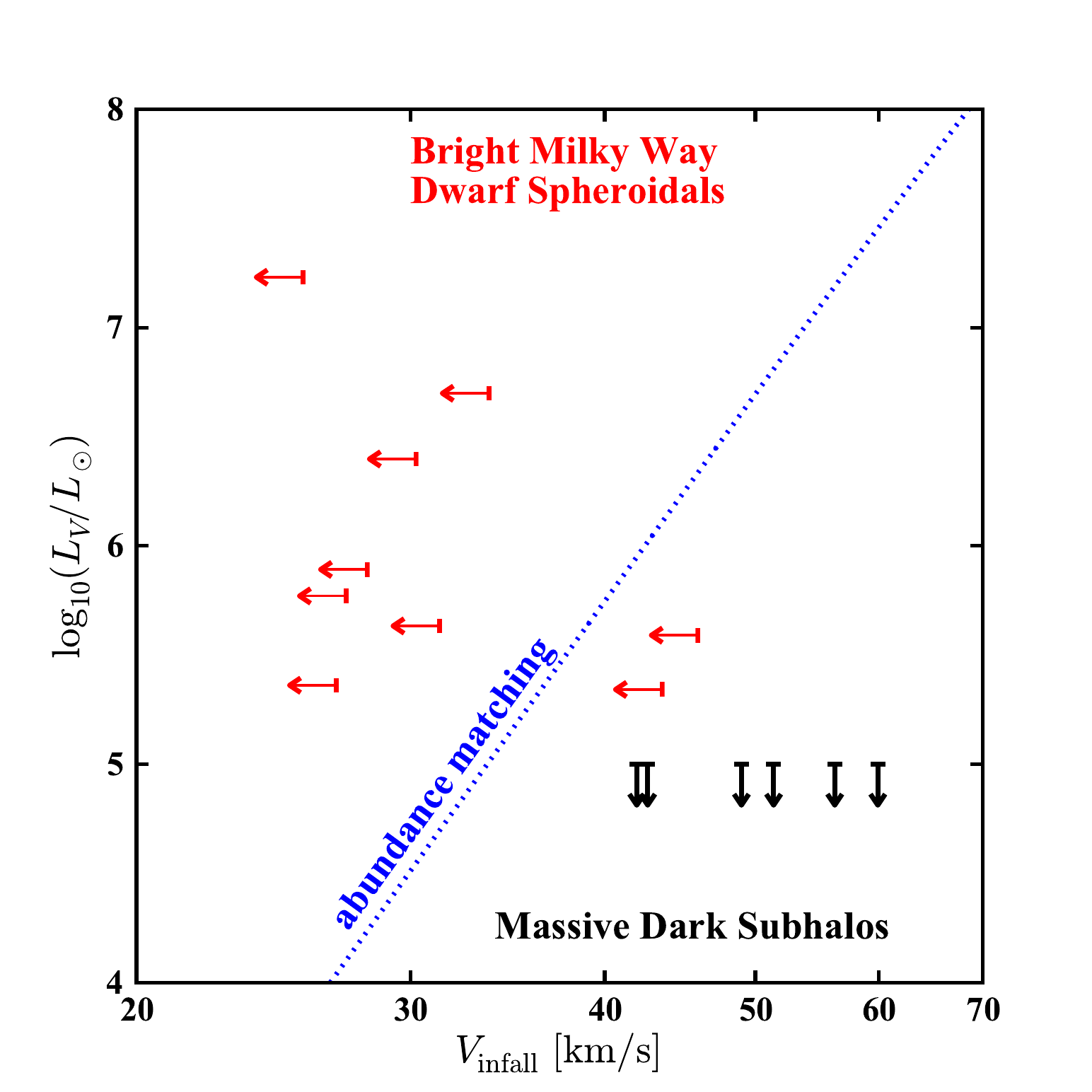}
 \caption{Relation between $\vacc$ and $L_V$ for Milky Way dwarf spheroidals
   (red points) and massive dark subhaloes (black points) for one representative
   halo realization (Aq-A).
\label{fig:sham}
}
\end{figure}

\begin{figure}
 \centering
 \includegraphics[scale=0.54, viewport=20 0 400 420]{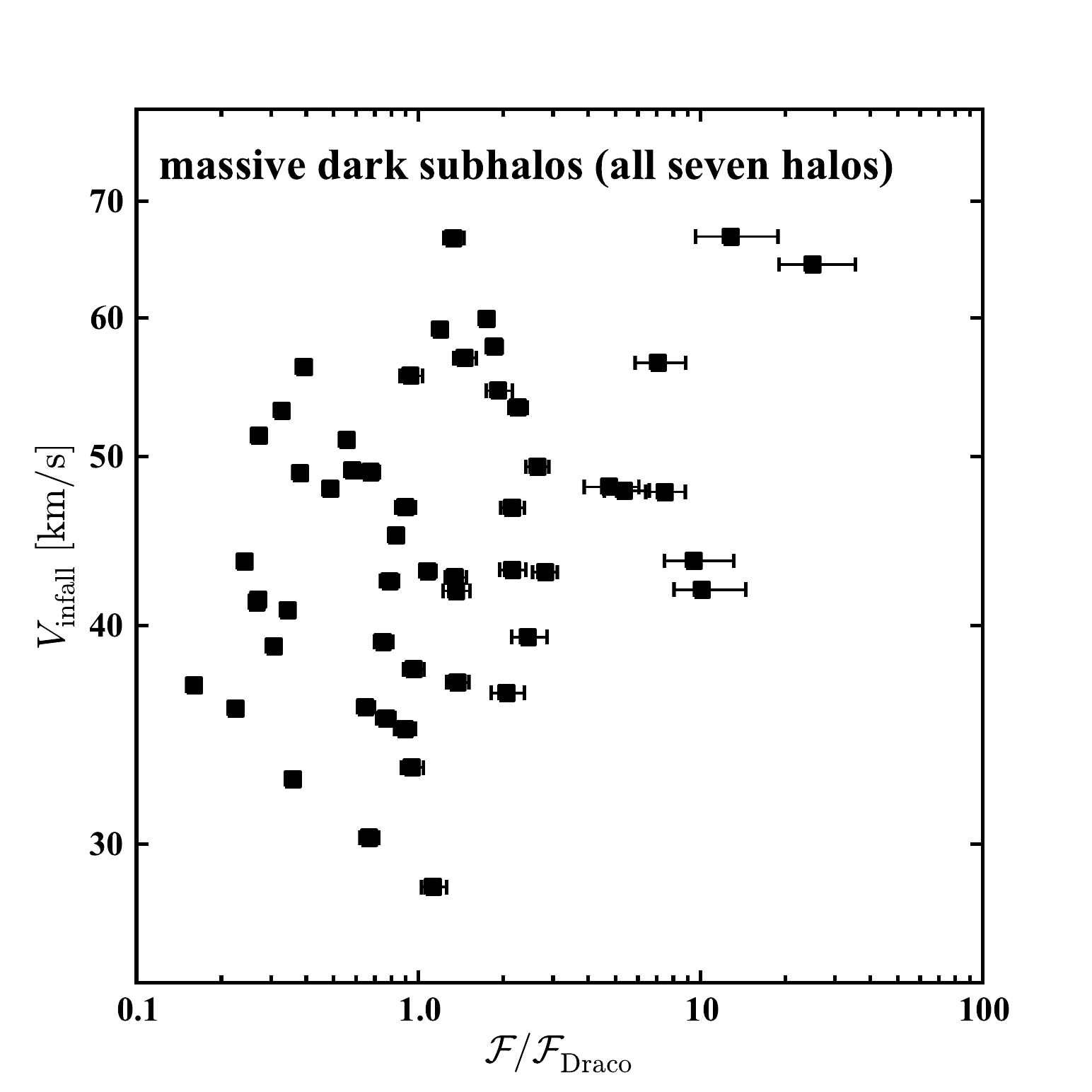}
 \caption{Distribution of annihilation fluxes from dark subhaloes, normalized
   to a typical scenario for the annihilation flux from Draco.
   Error bars reflect 68\% confidence levels for varying the specific angular location of
   the observer on the solar circle.  The typical halo has approximately four dark
   subhaloes with annihilation fluxes exceeding that of Draco. 
 \label{fig:ann_flux}
}
\end{figure}

If massive, dark subhaloes do exist in the Milky Way halo, their presence has
important implications for indirect dark matter searches.  Denser subhaloes
produce a larger luminosity from dark matter annihilation; from
Fig.~\ref{fig:rmax_vmax}, the dark subhaloes expected in the Milky Way are
denser than their potentially luminous counterparts and therefore may be bright
in $\gamma$-rays due to annihilations.  In Fig.~\ref{fig:ann_flux}, we plot the
annihilation flux, $\flux$, within $2.4 \times 10^{-4}\,$ steradians (a circular
region with radius 0.5 degrees) of the center of each massive dark subhalo
relative to the predicted flux within the same angular radius originating from
Draco, one of the most promising targets among the MW dwarfs for Fermi
\citep{abdo2010}. The horizontal error bars on the data points show 68\%
confidence intervals based on 1000 random realizations for the observer's
location (constrained to have a galactocentric distance of 8 kpc).  Dark
subhaloes are promising indirect detection candidates: each halo has at least
two dark subhaloes with annihilation fluxes larger than that of Draco, and four
of the seven haloes have at least one dark subhalo with $\flux/\flux_{\rm Draco}
>5$.

Our division between dark and bright subhaloes is very conservative: rather than
requiring at most one subhalo that is consistent with each of the bright MW
dwarf spheroidals, we require that {\it all} of the dark subhaloes are
inconsistent with {\it all} of the bright dwarfs at the $2\,\sigma$ level.
While the quantitative results may, in principle, change slightly if systematic
errors in the determination $\mhalf$ for the densest MW dwarfs (Draco and Ursa
Minor) have resulted in an underestimate of $\mhalf$, our general result -- that
many of the most massive simulated subhaloes are too dense to host any bright MW
satellite -- will be unaffected unless all of the measured $\mhalf$ values
change substantially.

\section{Discussion}
\label{section:discussion}
The results of Section~\ref{section:results} show that high resolution \lcdm\
simulations of individual galactic haloes generically predict that the Milky Way
should host at least six subhaloes that, at one point, had maximum circular
velocities in excess of 30 $\kms$ and yet are incompatible with any known MW
satellite (including the Magellanic Clouds) having $L_V > 10^{5}\,\lsun$.
Either these subhaloes actually exist as predicted in the Milky Way, requiring us
to understand their properties and stellar content, or they do not exist, in
which case we must understand the mechanism(s) that suppress their formation or
survival.

\vspace{0.1cm}
\noindent\textbf{If massive dark subhaloes exist as predicted}: \\
\indent 
  Detecting massive dark subhaloes would be a strong confirmation of the standard
  \lcdm\ paradigm.  These dark subhaloes might host at least some of the recently
  discovered ultra-faint galaxies, all of which have luminosities lower than
  $10^{5}\,\lsun$.  
  Kinematic constraints favor masses and densities for the ultra-faints that are
  indicative of fairly massive subhaloes \citep{strigari2008, walker2009,
    simon2010}, albeit with large uncertainties at present (e.g.,
  \citealt{wolf2010, martinez2010}).  If some of the ultra-faints are hosted by
  the massive subhaloes described here,
  they would have total mass-to-light ratios of $10^{5}-10^{8}$.  The
  ultra-faints would be excellent candidates for indirect dark matter detection
  in this scenario (Fig.~\ref{fig:ann_flux}).  An alternate detection method
  could be through the subhaloes' tidal influence on the MW's HI disk
  \citep{chakrabarti2011}. While the existence of effectively dark subhaloes
  with low masses is
  perhaps not surprising given the standard \lcdm\ power spectrum and the
  variety of effects that can impede cooling and star formation in shallow
  gravitational potential wells, the prospect of subhaloes more massive than the
  hosts of bright dwarf spheroidals but with $L_V < 10^{5}\,\lsun$ is intriguing.

  The existence of massive dark subhaloes requires that the fundamental
  assumption of abundance matching models -- that galaxy stellar mass or
  luminosity is a monotonic function of $\vacc$ -- does not hold for $\vacc \la
  50 \,\kms$; Fig.~\ref{fig:sham} illustrates this point.  Galaxy formation on
  scales below $50\,\kms$ should therefore be effectively stochastic, with
  stellar mass depending sensitively on specific details of a subhalo's
  environment, formation history, etc. rather than primarily determined by host
  halo mass or $\vacc$.
\vspace{0.1cm}

\noindent \textbf{If massive dark subhaloes do not exist as predicted:}\\
\indent The most prosaic explanation is that the haloes studied here are
  not representative of the MW-mass halo population at large in \lcdm.  This is
  unlikely to be the case for the Aquarius haloes, however, since they have
  substructure abundances typical of the full sample of over 2000 MW-mass haloes
  from the Millennium-II Simulation's $(137\,\mpc)^3$ volume
  \citep{boylan-kolchin2010}.  The subhalo mass function of the Milky Way could
  also be a statistical anomaly with respect to \lcdm\ expectations, in the
  sense that massive MW subhaloes are all less concentrated than expected, or
  that there are zero massive dark subhaloes in the MW when we expect at least
  six (Fig.~\ref{fig:vacc_cum}).  Perhaps the best way to investigate this
  possibility is to obtain detailed kinematic measurements of M31's satellites:
  if the M31 satellite system does not require massive dark subhaloes in \lcdm\
  simulations, then the statistical anomaly explanation would gain more
  traction.

 If the MW's subhalo mass function is not aberrant, then understanding why
  there are no massive dark subhaloes would likely result in important insight
  into the physics governing structure formation.  One possible \lcdm-based
  explanation is that the dark matter distribution in satellites of the MW is
  substantially less concentrated than current dissipationless simulations
  predict.\footnote{Lower concentrations of dark matter are also favored by many
    observations of low-mass field galaxies (e.g.,
    \citealt{kuzio-de-naray2008}), although these tend to be gas-rich,
    disk-dominated systems with higher luminosities than the bright MW dwarf
    spheroidals.}  Baryonic processes may affect the dark matter distribution on
  small scales by heating it to larger radii, which would have the desired
  effect of lowering the dark matter density.  For such a solution to work, it
  would have to substantially lower the dark matter density on scales of 0.3 --
  1 kpc (corresponding to the deprojected half-light radii of the bright dwarf
  spheroidals) while not strongly impacting the dark matter on smaller scales
  ($\la 100\,{\rm pc}$, corresponding to the half-light radii of the ultra-faint
  dwarfs), as ultra-faints seem to have high central dark matter densities
  \citep{simon2010}.  It is not clear that this could produce nearly identical
  average dark matter densities on scales of 300 pc in galaxies spanning a
  factor of $10^4$ in luminosity \citep{strigari2008}.  

  Gravitational shocks from encounters with the MW disk may also destroy some
  fraction of satellites \citep{donghia2010}.  This mechanism works most
  efficiently at destroying low-mass subhaloes, however.  Furthermore, it would
  not affect subhaloes that have larger pericenters or were accreted recently;
  many massive dark subhaloes in the simulations studied here fall into these two
  categories.

  If the Milky Way's dark matter subhalo population is typical of \lcdm\
  predictions, and baryonic physics has not strongly modified the internal
  structure or abundance of massive subhaloes, then the more drastic solution of
  modifying the underlying cosmology may be required in order to circumvent our
  primary conclusions that massive dark subhaloes should exist and that galaxy
  formation on small scales is stochastic.  Merely tweaking the cosmological
  parameters within the \lcdm\ model is unlikely to have an influence, as VL-II
  and Aquarius bracket current estimates of $\sigma_8$ and $n_s$.  Modifying
  the dark matter power spectrum on sub-galactic scales -- for example, through
  Warm Dark Matter (WDM) with a characteristic scale of 40 to 50 $\kms$ -- would
  result in both fewer massive subhaloes (e.g., \citealt{zavala2009}) and lower
  central densities in such subhaloes. Recent analyses of the Ly-$\alpha$ forest
  put fairly stringent constraints on the mass of WDM particles, however
  \citep{boyarsky2009}.  Dark matter self-interactions would also reduce the
  central densities of subhaloes, and would additionally make them more
  vulnerable to tidal disruption.  It is far from obvious that the abundance and
  dynamics of observed MW satellites would be correctly reproduced in the viable
  parameter space of these non-CDM models.

\vspace{0.2cm}

In summary, we find that the majority of the most massive subhaloes in
dissipationless \lcdm\ simulations are too dense to host any of the bright Milky
Way satellites.  It follows that galaxy formation must be effectively stochastic
in haloes with maximum circular velocities of $V \la 50 \,\kms$.  This conclusion
may be circumvented if the Milky Way's subhalo population differs substantially
from the average \lcdm\ expectation, or if the abundance or structure of massive
subhaloes in the Milky Way is strongly affected by baryonic processes or
different dark matter physics.

\section*{Acknowledgments} 
We thank Louis Strigari for interesting discussions and the Aquarius and Via
Lactea collaborations for providing access to their simulation data.  The
Aquarius Project is part of the program of the Virgo Consortium for cosmological
simulations.  The Millennium and Millennium-II simulation databases used in this
paper were constructed as part of the activities of the German Astrophysical
Virtual Observatory.  JSB was supported by NSF AST-1009973; MK was supported by
NASA grant NNX09AD09G.

\bibliography{draft}
\label{lastpage}
\end{document}